# Neutron shielding and activation of the MASTU device and surrounds

David Taylor[a], Steven Lilley[a], Andrew Turner[a], Andrew Davis[b]

[a]*EURATOM/CCFE Fusion Association, Culham Science Centre, Abingdon, OX14 3DB, UK*
[b]*Now at College of Engineering, University of Wisconsin, Madison, Wisconsin 53706, USA*

A significant functional upgrade is planned for the Mega Ampere Spherical Tokamak (MAST) device, located at Culham in the UK, including the implementation of a notably greater neutral beam injection power. This upgrade will cause the emission of a substantially increased intensity of neutron radiation for a substantially increased amount of time upon operation of the device. Existing shielding and activation precautions are shown to prove insufficient in some regards, and recommendations for improvements are made, including the following areas: shielding doors to MAST shielded facility enclosure (known as "the blockhouse"); north access tunnel; blockhouse roof; west cabling duct. In addition, some specific neutronic dose rate questions are addressed and answered; those discussed here relate to shielding penetrations and dose rate reflected from the air above the device ("skyshine"). It is shown that the alterations to shielding and area access reduce the dose rate in unrestricted areas from greater than 100 μSv/hour to less than 2 μSv/hour averaged over the working day.

The tools used for this analysis are the MCNP (Monte Carlo N-Particle) code, used to calculate the three-dimensional spatial distribution of neutron and photon dose rates in and around the device and its shields, and the nuclear inventory code FISPACT, run under the umbrella code MCR2S, used to calculate the time-dependent shutdown dose rate in the region of the device at several decay times.

## 1. Introduction

MAST is a Spherical Tokamak that has been operating at Culham since 1998 [1]. A project to upgrade MAST (to "MAST Upgrade" – MASTU) is underway with the modifications being performed in a shutdown planned from October 2013 to March 2015 [2]. The neutral beam heating power (currently ~3 MW) will initially be ~5MW delivered from the existing 2 injectors and will be progressively upgraded to 10MW by the addition of 2 more injectors. Following the upgrade the duration of the plasma will progressively be increased from the current maximum of ~1s to ~5s. Consequently the neutron yield from each tokamak pulse, which is predominantly from heating beam ion interactions with the target plasma (i.e. from Deuterium-Deuterium fusion, producing 2.45 MeV neutrons), is anticipated to increase by at least an order of magnitude for the longer pulses planned.

MAST is surrounded by an access-controlled concrete blockhouse for radiation shielding purposes. The device and the blockhouse have been modelled in an MCNP input file, illustrative cross-sections from which are shown in figure 1.

## 2. Shielding analysis

To make the calculations described below, the following tools were used. MCNP5 [3] has been used with the FENDL2.1 neutron cross-section library [4] to provide Monte Carlo simulations of neutron transport. FISPACT [5] is a zero-dimensional nuclear inventory code, developed at CCFE (formerly UKAEA). Taking as input an arbitrary material composition, neutron spectrum and irradiation schedule, it evolves a time-dependent inventory. MCR2S [6], also developed at CCFE, is a tool for the generation of activation gamma sources at selected times based on the rigorous-2-step formalism [7], MCNP 3D mesh-tallies, and the FISPACT code. The output of the code can be used by a modified MCNP source routine to perform transport to estimate quantities such as shutdown dose rate.

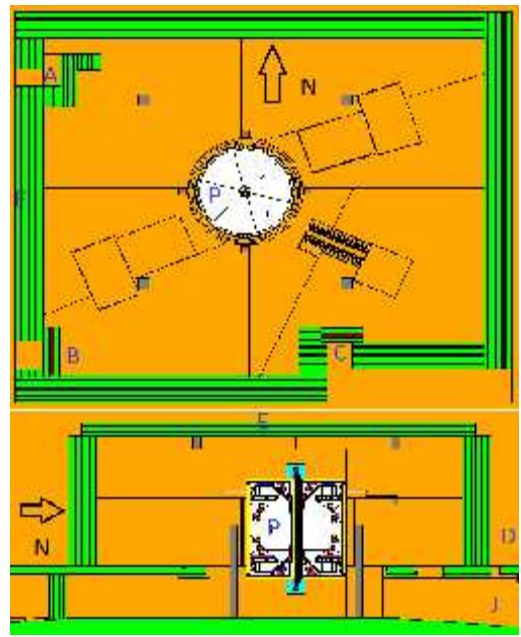

Fig. 1. MAST device (central) and blockhouse (surrounding concrete). The upper plot shows a midplane horizontal slice of the volume modelled (with North being upwards on the page) while the lower plot

---

*author's email: david.taylor@ccfe.ac.uk*

shows a South-North vertical slice. The MAST vessel is a circular cylinder with a diameter of 4 metres, while the interior of the blockhouse is approximately a square with side length 20 metres. A distance scale is shown on figure 2. Letters indicate referenced portions of the blockhouse: A – NW labyrinth; B – SW door; C – S door; D – North annexe / "Tiger Cage"; E – roof; F – W cabling duct; J – North access tunnel, P – MASTU vessel.

### 2.1 Onsite criterion

The MAST blockhouse lies within building D1 of the Culham site. Areas of this building near to the blockhouse are in use by non-classified radiation workers, and are subject to various dose rate limits, the strongest of which is a voluntary institution limit of 1 mSv/year. MASTU at its full extent is anticipated to run at a maximum pulse rate of 2 pulses per hour for an 8 hour working day on 126 days per year, with pulses lasting for a maximum of 5 seconds. Hence an operational ceiling of approximately 2000 pulses per year is established. However, the original MAST safety case [8] was produced on the assumption of 1000 full power pulses per year, and in practice this has proved highly conservative, with no annual MAST neutron budget approaching within a factor of 20 of this assumption. Bearing in mind that MAST is able to pulse at twice the rate of MASTU, continuing the conservative assumption of a maximum of 1000 full power shots per year into MASTU operation seems fully justified.

Thus we convert the voluntary dose rate limit of 1 mSv/year to an equivalent 2 μSv/hour (averaged over an 8 hour working day), a figure that we use as our dosage cut-off limit for uncontrolled areas. This is equivalent to an instantaneous dose rate limit of 0.72 mSv/hr during pulsing. Neutron fluence to dose conversion factors were taken from the ANSI-ANS-91 standard [9].

### 2.2 Limitations of pre-existing shielding

Replacing a MAST-strength source with a MASTU-strength source whilst retaining the pre-existing level of shielding reveals a number of regions in need of extra neutronic protection. Figure 2 shows dose rate maps for MASTU "Stage 2" emission (i.e. for the ultimate planned upgrade, using an on-load source rate of $1.6 \times 10^{15}$ n/s) corresponding to the regions shown in figure 1 plus North and South extensions; the dark blue region beyond the purple contour is the region of acceptable dose rate by the above criterion.

A number of weaknesses are readily apparent – i) The current NW labyrinth ("A" in figure 1) would permit an excessive amount of radiation to escape via the entrance; ii) The areas immediately beyond the blockhouse walls would see a large amount of radiation; iii) The SW ("B") and S ("C") shielding doors would be weak spots within this; iv) The Northern extension (known as the "tiger cage") ("D") would see a large dose emanating up the stairs from the basement; v) The current roof ("E") would permit a large dose to escape through it; and vi) A cabling duct to the West ("F") would allow some degree of neutron streaming.

To address these inadequacies, the following changes have been adopted – i) Closure of the Western corridor during operations ("K" in figure 3 – note that a large region West of this corridor is already closed during operations); ii) Closure of the South annexe ("L"); iii) Thickening of the East wall of the blockhouse ("M") from 120 cm to 160 cm; iv) Redesigning of the tiger cage area ("D"), including the building of a long Northern concrete wall ("N"); v) Thickening of the blockhouse roof from 60 cm to 80 cm; vi) Redesigning the NW labyrinth ("A"); vii) Creating of a labyrinth outside the South door ("O").

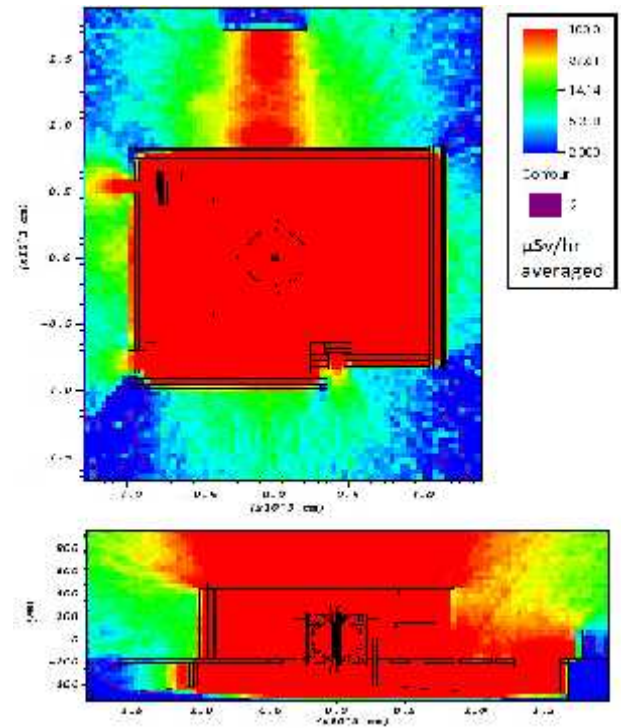

Fig. 2. Two dose rate maps for the existing MAST shielding, corresponding to the two plots shown in figure 1. Both plots are extended to North and South, which are respectively above and below in the upper plot and left and right in the lower plot. The extremes of the colour scale are dark blue (<2 μSv/hour averaged) and red (>100 μSv/hour averaged). The statistical error on the calculations is lower than 1% within the blockhouse, rising to 5-10% within the tiger cage and 15-20% in the regions beyond the shielding walls. The voxels in the mesh are of size 25 cm x 25 cm x 25 cm, as they also are in Figure 4.

### 2.3 Improvements to pre-existing shielding

An improved version of figure 1 is shown in figure 3, along with corresponding dose rate maps in figure 4. This illustration also contains several other alterations, discussed below.

The dose rate in non-restricted areas is now successfully ameliorated, noting that the substantially redesigned tiger cage area is restricted during operations. An additional neutron contribution to the on-load dose rate in this area is created by the need to drill a straight penetration through the North blockhouse wall for the operation of the CO2 laser diagnostic. With the proposed polythene backstop for this penetration ("G" in figure 3), the dose in this area is still dominated by the dose emitted via the MASTU basement.

Another clear structural change is seen in the NW labyrinth ("A"), in which the shielding has been both lengthened and thickened. Various other options were considered, but this was found to be a solution that reduced both the direct dose through the wall and the indirect streaming dose around the labyrinth to acceptable levels whilst having minimal negative impact on the space available inside the blockhouse for tokamak diagnostics. Other illustrated structural changes include the thickened East wall ("M") and the new labyrinth outside the South door ("O").

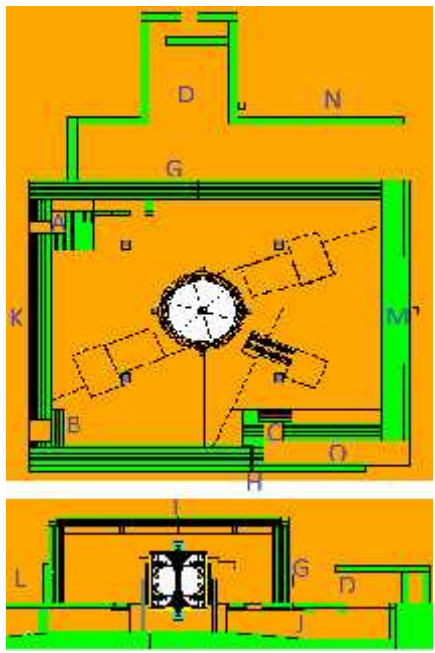

Fig. 3. Layout of facility incorporating later refinements. Slices and A, B, C, D, J as in figure 1. G is the polythene backstop to the CO2 laser penetration, H is the gas pipe penetration, I is the roof flange penetration, K is the Western corridor, L is the South annexe, M is the East blockhouse wall, N is the Northern concrete wall, O is the South door labyrinth.

**2.4 Blockhouse penetrations**

In addition to the Northern blockhouse penetration for the CO2 laser described above, a number of other blockhouse penetrations are required. The first of these is shown in figure 3, marked "H". This penetration is designed to permit the passage of gas pipes into the blockhouse, and contains a labyrinth to reduce direct streaming. It may be seen from figure 4 that the natural shielding provided by the existing concrete structure to this type of penetration is most effective.

The second involves the replacement of a small circular section of the blockhouse roof directly above the centre of the tokamak ("I") with a steel flange through which pass three small tubes in vacuum, through which pellets may be injected into the plasma. In the absence of this change, the roof, a restricted region of high dosage, sees a local dose rate minimum directly above the central column of the tokamak, due to the toroidal nature of the neutron source. The insertion of the flange into the model does not qualitatively change this dose rate distribution, and the change in the peak of the roof-top distribution is by an amount smaller than the margin of statistical error.

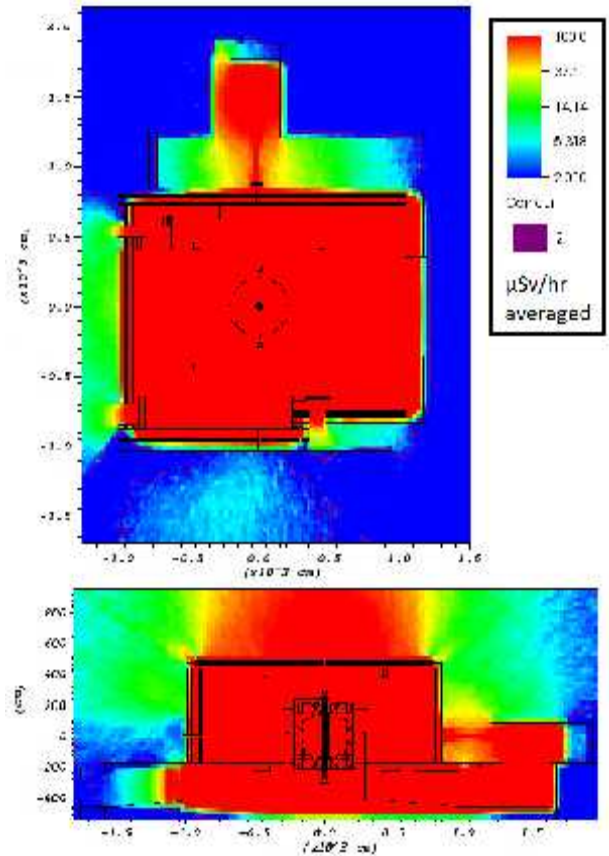

Fig. 4. Dose rate maps over facility using the layout shown in figure 3. Slices and colour scale as in figure 2. The statistical error on the calculations is lower than 1% within the blockhouse, rising to 5-10% within the tiger cage and 15-20% in the regions beyond the shielding walls.

The third involves the drilling of a straight penetration in the South wall large enough to accommodate 1,100 cables attached to magnetic diagnostics. It was found that angling a cylindrical penetration NW-SE did not significantly increase the dose rate external to the blockhouse.

**2.5 Skyshine effects**

It will be necessary from time to time to remove the roof of the blockhouse in order to take out or insert large items. The hazards of operation with an improperly replaced roof were analysed. Four cases were considered

– i) Reference case, with 80cm thick blockhouse roof; ii) Completely absent roof; iii) One roof layer missing, with 60cm thick roof; and iv) One roof tile missing, a long penetration through the roof, of dimension 0.978m x 8.65m.

A conservative overestimate of the hazard to the general public in each case is obtained by calculating the dose rate at the distance that is a little lower than the real minimum distance from the MASTU vessel to the Culham site boundary – 200m; the real distance is approximately 240m. The inherent difficulty in obtaining convincing particle statistics at this remote location is overcome with the use of a dxtran sphere [10] placed around the relevant particle tally. In line with regulation 134 of [11], the acceptable public dose rate is taken to be 0.3 mSv/year, equivalent to 0.6 µSv/hour averaged.

Using this, we calculate that the site boundary dose rate for MASTU "Stage 2" when the full roof is in place is a mere $9.19 \times 10^{-4}$ µSv/hour (statistical error 13.4%), averaged over an 8-hour working day as before, which is a tiny fraction of the permissible dose rate. Reducing the roof by a single layer to the current 60cm thickness increases this dose rate to a still negligible $8.56 \times 10^{-3}$ µSv/hour averaged (error 14.2%). Omitting a single roof tile through both layers increases the dose rate further, to 0.685 µSv/hour averaged (error 6.81%), a small amount over the limit. Of these tested roof layouts, only the worst case scenario of complete roof omission results in a dose rate significantly above the public limit; this scenario results in a comparative boundary dose rate of 15.9 µSv/hour averaged (error 4.95%). For these calculations, an F4 MCNP tally was used, averaging over a cube of side 200 cm.

Dose rates for these cases were also calculated using mesh tallies for a region on the ground floor of the larger building in which the blockhouse is contained. In the absence of a known precise location for the upgraded MASTU control room, an unshielded region between 8 and 16 metres South and 30 and 40 metres East of the MASTU central point was used, a space abutting the Eastern edge of the D1 building. Once a precise location is decided upon, this analysis will be rerun, but the conclusions are not anticipated to change. For the reference case of full designed shielding, this region sees a maximum dose rate of approximately 0.5 µSv/hour averaged, well inside the specified limit given in section 2.1. In contrast, the omitted roof case sees a maximum dose rate of approximately 1 mSv/hour averaged – a long way over the limit, but also a dose rate unlikely to cause physical harm unless sustained for a long time. In the case where one layer of the roof is omitted completely, the "control room" sees dose rates up to approximately 2 µSv/hour averaged – still marginally within the unrestricted limit, but the case where there is a long penetration in the blockhouse roof presents a higher dose rate, of up to 30 µSv/hour averaged.

## 3. Activation analysis

A series of activation timesteps were calculated following a worst case set of assumptions for MASTU operation. It was assumed that the time of interest immediately follows the cessation of MASTU operation on the completion of the project after the currently anticipated 10 years. In line with current planning, each year was taken to consist of 126 experimental days, with each experimental day consisting of 8 hours of operation, and each hour containing two 5 second pulses, emitting at full anticipated power for the device throughout the pulse. This is evidently a much more intense irradiation schedule than will actually come to pass; a similar set of assumptions made for the original MAST safety case predicted an annual neutron budget of greater than 20 times the highest recorded annual emission.

The calculations were made using MCR2S, which couples MCNP with FISPACT. Neutron spectra were computed by MCNP on a 3D mesh. These were then passed into multiple FISPACT runs, which calculated a nuclear inventory for each mesh voxel, using the EAF 2007 data library [12]. Having thus simulated the spatially distributed material activation, rerunning MCNP in photon mode returns the shutdown dose rate map for each timestep.

The intention in running this analysis is to provide answers to a number of specific questions dealing with access to restricted areas. The three areas in question are the tiger cage region ("D"), the interior of the blockhouse, and the interior of the MASTU vessel. The tiger cage is of these the region of lowest shutdown dose rate by a large margin. Immediately after the cessation of the given operating schedule the dose rate does not exceed 0.1 µSv/hour in any part of this area. In consequence, it will be possible to access this area between pulses.

The interior of the blockhouse sees higher dose rates, and the principal source of activated gamma radiation within it is the MASTU vessel itself. Although areas of the blockhouse far from the vessel appear safe to access straight after operation, areas in the vicinity of the vessel require 'cooling-off' time, with the stainless steel of the vessel centre column displaying the greatest activity. For the given pulsing schedule the dose rate in the region immediately by the vessel requires substantial time to drop below the limit given in section 2.1 – many weeks. Plots of dose rate at selected times post-operation are shown in figure 5, which illustrate that there are at least two timescales of interest, dealing respectively with short-lived and long-lived activated isotopes. Entry to the blockhouse depends on the former, while entry to the vessel depends on the latter.

It is apparent that controlled access to the blockhouse will be possible within a few days after shutdown, even after this deliberately extreme pulsing schedule. It is also apparent that unrestricted blockhouse access will not be possible on a reasonable timescale given these assumptions. This interim conclusion should be seen to imply the need for more detailed and realistic modelling rather than as a limiting factor for maintenance; the current modelling is somewhat coarse in both mesh and cooling steps. Once secondary modelling is coupled with detailed maintenance requirements, a detailed task-based dose rate assessment can be made.

Regarding vessel access, the maximum dose rate here calculated within the vessel using mesh tallies is 40 µSv/hour after 11.57 days, 7.8 µSv/hour after 1 year, and less than the unrestricted limit after 3 years.

dose rate in unrestricted areas now brought below the derived limit of 2 µSv/hour.

Sufficient shutdown dose rate calculations have been made to provide a general overview of access constraints during maintenance. The next phase of work on this project will provide more detailed analysis of this subject once maintenance requirements are finalised.

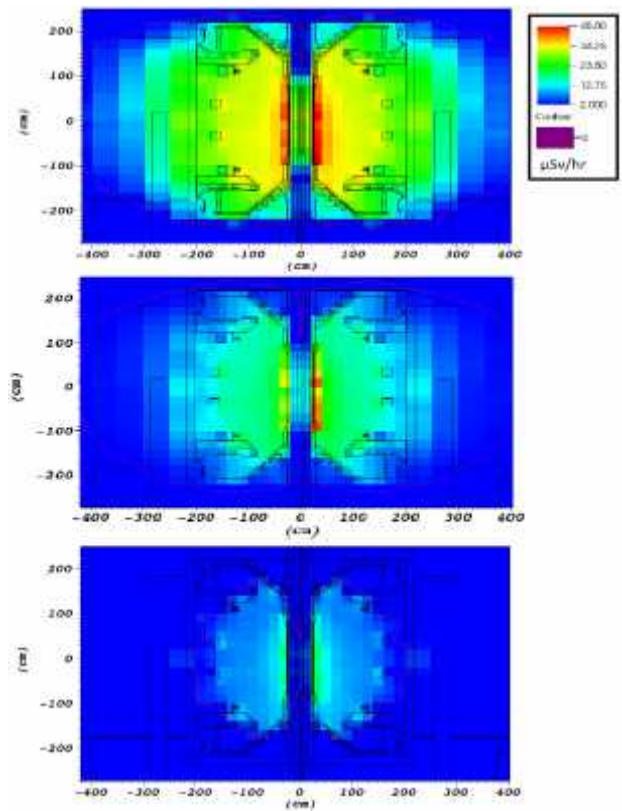

Fig. 5. Dose rate due to gamma activation in the region of the MASTU vessel for three selected timesteps post-operation: 60 µs (top), $10^6$ s (11.57 days) (middle), and 1 year (bottom). The pictured section shows the MASTU vessel and immediately adjacent areas in a South-North section extending 4m either side of the MASTU central point and 2.5m each way vertically. The extremes of the colour scale are dark blue (<2 µSv/hour) and red (>45 µSv/hour). The voxels in the mesh are of size 20 cm x 20 cm x 10 cm.

## 4. Conclusions

The existing shielding for the MAST vessel has been analysed allowing for the planned experimental upgrades to the tokamak. Recommendations have been made on improvements to this shielding, and the modelling of on-load radiation dose rate indicates that these recommendations are effective. This covers a number of analyses in a complex 3D geometry, dealing with general questions of bulk shielding, penetrations, and skyshine in addition to a number of specific points. Figures 2 and 4 show dose rate maps before and after shielding improvements, with the working day averaged

## Acknowledgements

This work was funded by the RCUK Energy Programme under grant EP/I501045 and the European Communities under the contract of Association between EURATOM and CCFE. The views and opinions expressed herein do not necessarily reflect those of the European Commission.